\documentclass{aa}  

\usepackage{bm}
\usepackage{graphicx}
\usepackage{txfonts}
\usepackage{xcolor}
\usepackage{hyperref} 
\newcommand*{\pd}{\partial}

\begin{document} 
    \title{Along the Primary Curve: Simultaneous Source and Lens Reconstruction of Bright Arcs in Cluster Lenses}
    
    \author{Atınç Çağan Şengül
          }

    \institute{Pittsburgh Particle Physics, Astrophysics, and Cosmology Center (PITT PACC),\\ 
    Department of Physics and Astronomy, University of Pittsburgh, \\
    3941 O'Hara Street,
    Pittsburgh, PA, 15260\\
    \email{aa2@pitt.edu}
    }

 
  \abstract
   {Gravitational lensing is the phenomenon arising when light rays are deflected by the mass between the source and the observer. Largely magnified and highly distorted images of background galaxies are formed by these angular deflections if the deflecting mass distribution and the background sources align. As the most massive gravitationally bound objects in the universe, galaxy clusters are places where such alignments are usually found. By carefully analyzing the images of lensed galaxies, one can measure the mass, both visible and invisible, along the line-of-sight. These measurements are crucial in investigating the nature of dark matter, which constitutes most of the mass within clusters. }
   {Existing lensing analysis methods typically forward model the multiple images of dozens of background galaxies lensed by the cluster. To make this forward modeling computationally tractable, these multiple images are reduced to a much smaller summary data vector, which includes the locations, magnifications, and distortions. Our work avoids this loss of information by forward modeling the data at pixel-level.}
   {We develop a parametric model for the angular deflections near the bright arcs that allows us to control the shape of the curve that gives the directions of the eigenvectors of the Jacobian of the lensing matrix. The bright and extended images often follow such curves.}
   {We apply our analysis method to the bright arcs in gravitational lenses SDSS J1110+6459 and SDSS J0004-0103. We present our lens and source reconstructions for each system. With the application of our new method to many other lensing systems, we anticipate significant improvement in lens modeling near the critical curve, which will provide higher precision mass reconstructions for the deflectors. High-precision lens models allow for more robust de-lensing, which aids the studies of various highly magnified sources.
   }
   {}


   \maketitle
%

\section{Introduction}
Galaxy clusters (clusters from now on) are the largest gravitationally bound structures in the universe, containing hundreds (sometimes thousands) of galaxies along with vast amounts of hot intracluster gas and dark matter. Clusters form in regions with much higher than average density in the early universe, making them crucial for understanding the large-scale cosmic architecture.

A significant portion of a cluster's mass comes in the form of dark matter. Studying the distribution and effects of dark matter in galaxy clusters helps us learn about its properties. As dark matter does not emit, absorb, or scatter light in any measurable amount, we can only quantify its distribution via its gravitational influence. One such influence is gravitational lensing, which occurs when a massive object bends the light from a more distant object behind it. The mass of the cluster acts like a lens, distorting, magnifying, and sometimes creating multiple images of the background object. By analyzing the positions and shapes of these arcs and images, we can infer the distribution of mass that is causing the lensing.

Traditional approaches to modeling cluster mass typically involve identifying the positions of multiple images of lensed sources, calculating where these images map onto the source plane using the current lens model, and ensuring that the multiple images of the same source converge to the same point on this plane \citep{frontier_fields_cluster_lensing}. This process takes the full telescope image of the cluster and reduces it to a summary data vector, which is much faster to fit. In a typical cluster lens, tens to hundreds of multiply imaged sources may be present, providing useful constraints for the lens model and making modeling feasible, especially when additional observational information is used from stellar kinematics \citep{mona_kine17,bergamini_kine19}. Nevertheless, this data reduction results in a significant loss of information. These steps are deemed necessary because the full image of a cluster lens is typically too large to analyze at the pixel level, often containing millions of pixels, which makes the ray tracing calculations computationally costly. Additionally, global cluster lens models are often not worth fitting the pixel-level data as they can have root-mean-square values rarely less than $0.1''$ and sometimes as high as $0.5''$ for the differences between the model predicted and observed positions \citep{cluster_rms2, cluster_rms}, failing to reach sub-pixel precision. To address this challenge, our work focuses on multiple images of a single source that is stretched into a lensed arc, with the goal of building a lens model that can precisely fit the angular deflections near the arc.

Critical curves of cluster lenses provide the highest magnification regions in the sky. This powerful magnification boosts the apparent luminosity of distant objects that would otherwise be difficult to detect. There are several ongoing science cases, mainly involving the study of these highly magnified objects, that would benefit from precise modeling of extended arcs in clusters: caustic crossings of individual stars \citep{caustic_star_hst1, caustic_star_hst2,caustic_star_hst3,caustic_star_jwst1,caustic_star_jwst2,caustic_star_jwst3,caustic_star_jwst4,caustic_star_jwst5,caustic_star_jwst6}, gravitationally lensed supernovae \citep{lensed_supernova1,lensed_supernova2,lensed_supernova3,lensed_supernova4,lensed_supernova5}, and star clusters \citep{lensed_star_cluster1}. Having more precise lens models near the critical curve is crucial in making robust inferences on the source properties in all of these cases, leading to a better understanding of the nature of these highly magnified objects.

A new approach to modeling cluster lenses at pixel-level was previously employed by \cite{sengul_cab}, where the multiple images of a source galaxy lensed by a cluster are analyzed by a local lens model called \textit{curved arc basis} (CAB) first proposed by \cite{curved_arc_basis}. The main limitation of CAB is that as an approximation, it fails if the magnified arc covers a large extended region or when the angular deflections change significantly over a short distance. Therefore, it is suitable for modeling sets of multiple images of a source that are either far from the critical curve or small in extent. For this reason, the magnifications of the images that are suitable for the application of the method in \cite{sengul_cab} are significantly lower than those of arcs overlapping the critical curves. The lens model developed in this work is built to be valid across the entire lensed arc as the image crosses the critical curve, even multiple times. Our full model, consisting of both a lens model and a free-form source light model, is optimized to reproduce the image of the extended arc at the pixel-level, retaining all available information and avoiding the data loss inherent in previous methods. However, this approach does come with the trade-off that our lens model, by construction, is only valid in the vicinity of the lensed arc under analysis.

This paper is organized as follows: In \S \ref{sec:methods} we describe the mathematical formulation of our lens and source model, in \S \ref{sec:results} we show our results when we apply our model to various lensed arcs in clusters, and in \S \ref{sec:discussion} we discuss the implications and potential applications of our approach. The code used for our analysis is available at \url{https://github.com/acagansengul/alongprimarycurve}.

\section{Methods}\label{sec:methods}
In this section we will describe how we simultaneously model the angular deflections and the source light morphology to reconstruct the images of lensed arcs.

\subsection{Eigenvectors of the Lensing Jacobian}
\begin{figure}
    \centering
    \includegraphics[width=0.5\textwidth]{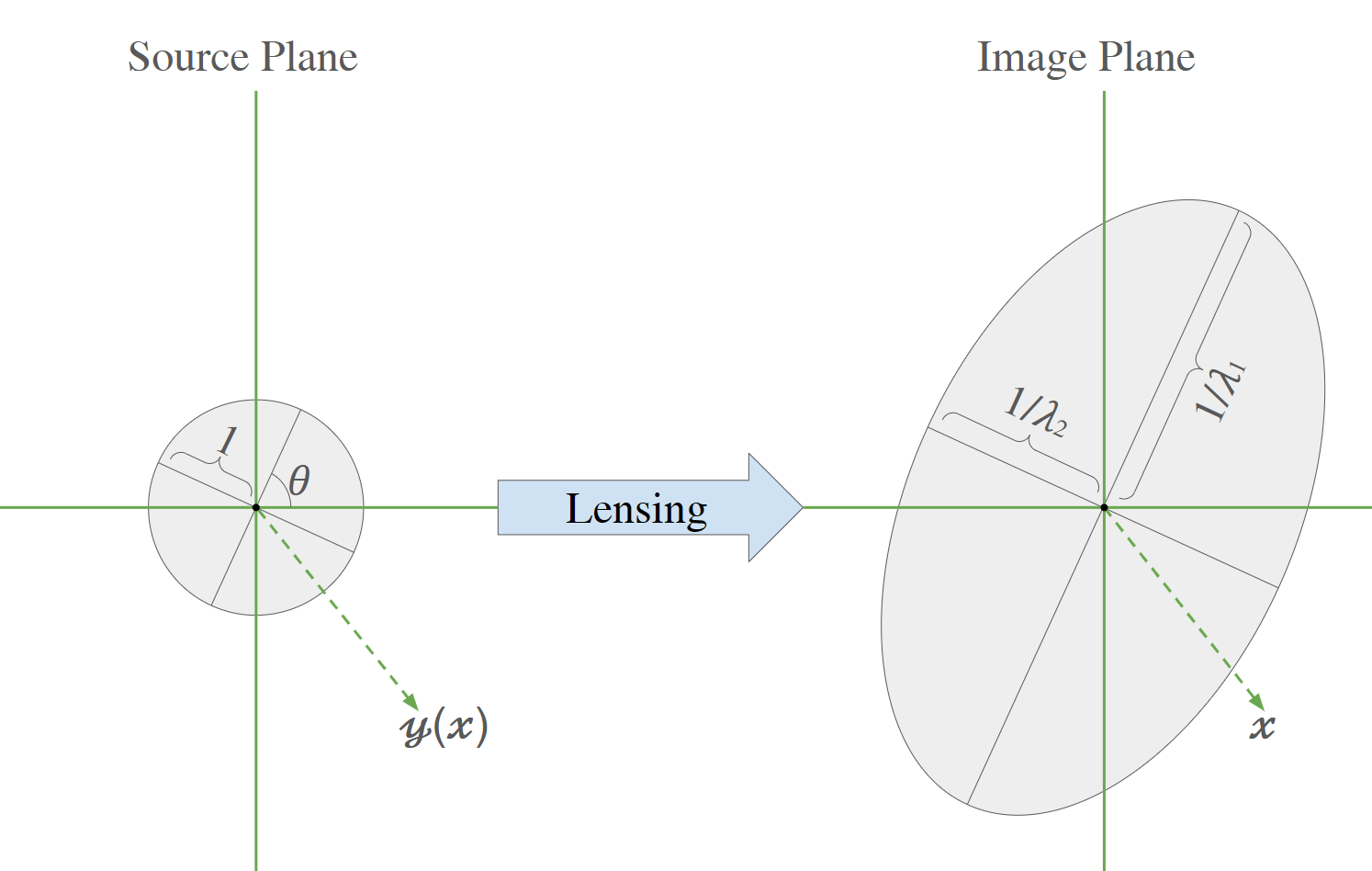}
    \caption{The effect of lensing on an infinitesimally small source. Locally a source that is circularly symmetric gets lensed into an ellipse. The eigenvectors of the Jacobian of the lensing function $\bm y$ (shown in Eq. \eqref{eq:jacob2}) are paralel to the semi-minor and semi-major axes of this ellipse. If the source has unit radius, the semi-minor and semi-major axes have lengths given by the eigenvalues of the Jacobian.}
    \label{fig:illus1}
\end{figure}

We can think of the source as a 2-dimensional brightness distribution in the background unlensed sky, which we will call the \textit{source plane}. The resulting image after lensing is also a 2-dimensional brightness distribution in the sky, which we will call the \textit{image plane}. The characteristic sizes of these images in the sky are rather small (arcsec scales) which allows us to use the flat sky approximation. Therefore, the angular position on the image plane can be mathematically represented as $\bm{x}\in \mathbb{R}^2$. The corresponding angular position on the source plane is given by the lensing function $\bm{y}(\bm{x})$ where $\bm y : \mathbb{R}^2 \rightarrow \mathbb{R}^2$. These are related by the lens equation:
\begin{equation}\label{eq:lenseq}
    \bm y (\bm x) = \bm x - \bm \alpha (\bm x)
\end{equation}
where $\bm \alpha$ is the angular deflection which is the result of the gravity of the mass distribution in the sky between the source and the observer. The Jacobian represents the distortion and magnification of infinitesimally small images:
\begin{equation}\label{eq:jacob1}
    \frac{\pd \bm{y}}{\pd \bm{x}} =
    \begin{pmatrix}
        \dfrac{\pd y_1}{\pd x_1} & \dfrac{\pd y_1}{\pd x_2}\\
        & \\
        \dfrac{\pd y_2}{\pd x_1} & \dfrac{\pd y_2}{\pd x_2}
    \end{pmatrix}.
\end{equation}
The eigenvalues and the eigenvectors of the Jacobian of the lensing function carries crucial information in understanding the magnification, distortion, and the orientation of multiple images that are formed by lensing \citep{grav_lens_textbook}. In the case of strong lensing, the dominant contribution to angular deflections comes from a galaxy or a galaxy cluster whose extent in the line-of-sight direction ($\sim 1\,\mathrm{Mpc}$) is much smaller than the cosmological distances ($\sim 10^3\,\mathrm{Mpc}$) between the source and the observer. This allows us to assume single-plane lensing where the lensing is done by a thin lens at a single redshift. The angular deflections of a thin lens have a symmetric Jacobian, because they can be written as the gradient of a scalar potential. A symmetric real matrix can be diagonalized with a rotation, as its eigenvectors are orthogonal:
\begin{equation}\label{eq:jacob2}
    \frac{\pd \bm{y}}{\pd \bm{x}} = \begin{pmatrix}
        \cos \theta & - \sin \theta \\
        \sin \theta & \cos \theta 
    \end{pmatrix}
    \begin{pmatrix}
        \lambda_1 & 0\\
        0& \lambda_2 
    \end{pmatrix}
    \begin{pmatrix}
        \cos \theta & \sin \theta \\
        -\sin \theta & \cos \theta 
    \end{pmatrix}.
\end{equation}
where we can choose $|\lambda_1| < |\lambda_2|$ without loss of generality. The geometric meaning of the components of Eq. \eqref{eq:jacob2} can be seen in Fig. \ref{fig:illus1}, which shows the effect of lensing on an infinitesimally small circular source with unit radius. The angle $\theta$ gives the orientation of the eigenvector with the smaller eigenvalue $\lambda_1$, which we will call the \textit{primary eigenvector}. 

The direction of the primary eigenvector tells us the direction along which the stretching of the source image due to lensing is maximum. If we "move" a point on the image along the direction of the primary eigenvector, the corresponding point on the source plane will move minimally. The direction in which the heavily magnified arcs appear to be stretched will typically lie along the direction of the primary eigenvector, assuming that the shape of the source galaxy is not unusually elongated. We make use of this property in the way parametrize our model for angular deflection field $\bm y$.

\subsection{Forming Multiple Images}

\begin{figure*}[h]
    \centering
    \includegraphics[width=0.75\textwidth]{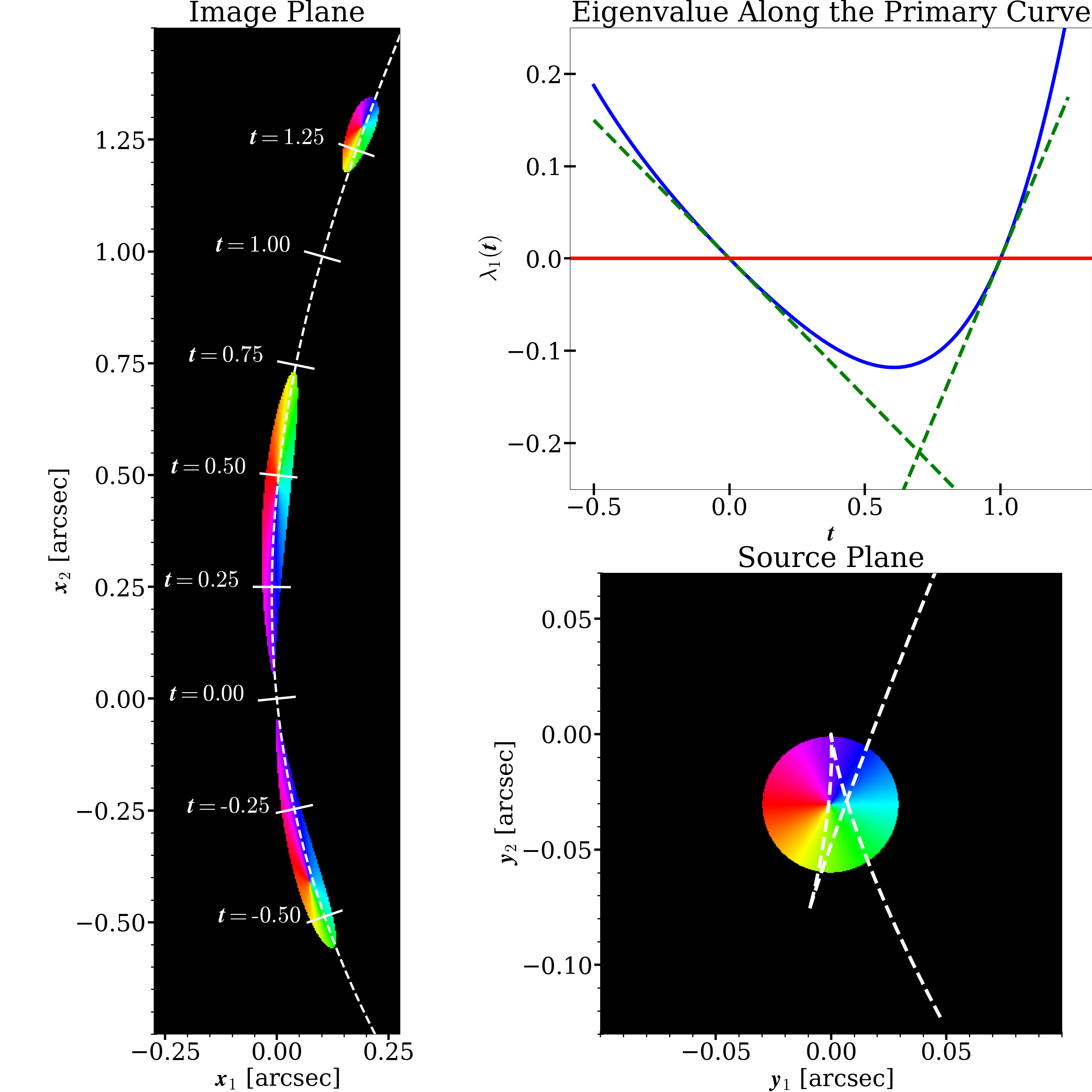}
    \caption{Qualitative illustration of lensing along a primary curve. \textit{Left:} 
    The lensed images of a disk shaped colorwheel source. The dashed curve shows the primary curve $\bm \xi(t)$ parametrized by $t$. The values of $t$ along the primary curve are shown as solid ticks.
    \textit{Top right:}
    The values of the primary eigenvalue $\lambda_1(t)$ along the primary curve. The places where $\lambda_1(t)$ hits 0 corresponds to the critical curve crossings where the magnification is infinite.
    \textit{Bottom right:}
    The unlensed image of the colorwheel source. The dashed curve shows the set of points that we get when we map the primary curve onto the source plane.}
    \label{fig:illus3}
\end{figure*}

Consider a curve $\bm \xi(t):\mathbb{R} \rightarrow \mathbb{R}^2$, which maps the coordinate $t$ to its corresponding point on the image plane, such that at each point this curve is tangent to the primary eigenvector of the lensing function. We will call any curve with this property \textit{primary curve}. We qualitatively illustrate  how images lie along the primary curve in Fig. \ref{fig:illus3} where we show the lensed image of a disk shaped source. The source is a colorwheel, shown in the \textit{bottom right} subplot, to make the effect of lensing clearer to see by eye. The dashed curve on the \textit{left} subplot shows the primary curve which is parametrized by $t$. The dashed curve in the \textit{bottom right} subplot is the curve that you get when you map the primary curve onto the source plane. The blue curve in the \textit{top right} subplot shows the eigenvalue $\lambda_1(t)$ as a function of the parameter $t$. If we start from the bottom of the primary curve and move along the curve with increasing $t$, we notice that initially $\lambda_1$ is positive. This means that on the source plane we also move along the same direction. As we keep moving and cross $t=0$, $\lambda_1$ flips to negative. As this happens, the movement on the source plane slows down, stops, and reverses direction. This forms the first kink of the dashed curve on the source plane. Because the movement on the source plane reversed directions, it goes over the source again to form the second image which is between $t=0.00$ and $t=0.75$. As we cross $t=1.00$, $\lambda_1$ flips back to positive resulting in another change in direction on the source plane. This forms the other kink of the dashed curve on the source plane. We cross over the source once more to form the third image which is around $t=1.25$. We can see from the colorwheel pattern that the parities of the first and third image are positive and the parity of the second image is negative.

\subsection{Parametrization of the Primary Curve}

\begin{figure*}[h]
    \centering\includegraphics[width=0.70\textwidth]{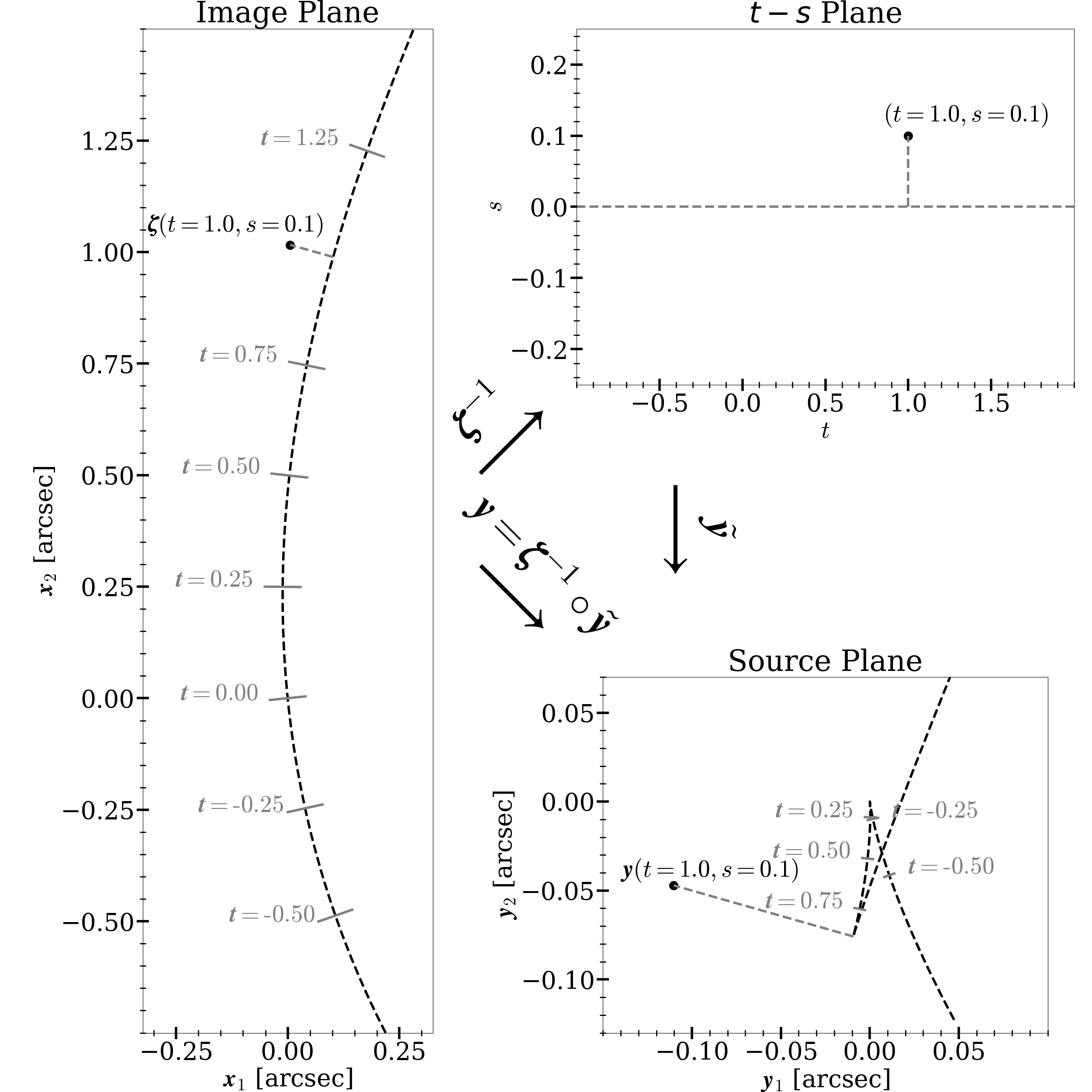}
    \caption{The flow of the lens model: The dashed line is the primary curve $\bm xi$ whose parametrization is shown in Eq. \eqref{eq:xidef}. Given a primary curve the function $\bm \zeta$ (shown in Eq. \eqref{eq:x_equation}) is a coordinate system that labels the points on the image plane with coordinates $t,s$. The transitive lensing function $\tilde{\bm{y}}$ (defined in Eq. \eqref{eq:transitive}) maps the points on the t-s plane onto the source plane.  Combining $\tilde{\bm{y}}$ with the inverse function $\bm \zeta^{-1}$ gives us the lensing function $\bm y$ that maps the image plane to the source plane.}
    \label{fig:illus2}
\end{figure*}

One can first parametrize a lensing function and calculate its primary curves. However, in this work, we first parametrize a curve and demand that this curve is a primary curve of our lensing function.

We parametrize the primary curve with $n+3$ parameters $(d_1,d_2,\theta_0,a_1,...,a_n)$. As determined by these parameters, the curve is given by
\begin{align}\label{eq:xidef}
    \bm \xi(t) = 
    \begin{pmatrix}
        d_1\\
        d_2
    \end{pmatrix}
    +
    Q_n(t)
    \begin{pmatrix}
        \cos\theta_0\\
        \sin\theta_0
    \end{pmatrix}
    +
    P_n(t)
    \begin{pmatrix}
        -\sin\theta_0\\
        \cos\theta_0
    \end{pmatrix}
\end{align}
where
\begin{align}
    Q_n(t) &= \int^{t}_0 dt'\,\cos\theta_n(t') \\
    P_n(t) &= \int^{t}_0 dt'\,\sin\theta_n(t')
\end{align}
and
\begin{equation}
    \theta_n(t) = a_1 t + a_2 t^2 + ... + a_n t^n.
\end{equation}
We choose a coordinate basis for the neighborhood of the primary curve with the function $\bm \zeta: \mathbb{R}^2 \rightarrow \mathbb{R}^2$. Motivated by the fact that the eigenvectors of the lensing function are orthogonal, we choose
\begin{equation}
    \bm \zeta \left[
    \begin{pmatrix}
    t\\
    s
    \end{pmatrix}\right]
    =
    \bm \xi (t)
    +
    \begin{pmatrix}
        0 &-1\\
        1 &0
    \end{pmatrix}
    \left(\frac{d\bm \xi}{dt}\right)s.
\end{equation}
This function $\zeta$ takes in a pair of coordinates $(t,s)$ and maps them to the corresponding point on the image plane. The first coordinate $t$ describes how far along the primary curve the point is while the second coordinate $s$ describes how far from the curve the point is. This is a bijective function (whose inverse we calculate in Appendix \ref{sec:quick_inv}) as long as $s$ is smaller than the curvature radius of $\bm \xi$. Using Eq. \eqref{eq:xidef}, we can obtain an explicit expression:
\begin{align}\label{eq:x_equation}
    \bm \zeta \left[
    \begin{pmatrix}
    t\\
    s
    \end{pmatrix}\right]
    =
    \begin{pmatrix}
        d_1 \\
        d_2
    \end{pmatrix}
    &+
    \left(Q_n(t) - s \frac{d P_n}{dt}(t)\right)
    \begin{pmatrix}
        \cos\theta_0\\
        \sin\theta_0
    \end{pmatrix} \\
    &+
    \left(P_n(t) + s\frac{dQ_n}{dt}(t)\right)
    \begin{pmatrix}
        -\sin\theta_0\\
        \cos\theta_0
    \end{pmatrix}.
    \nonumber
\end{align}

\subsection{Parametrization of the Lens Model}

We define the \textit{transitive lensing function} $\tilde{\bm y}(\bm u)$ as a function of $\bm u =   \begin{pmatrix}
        t\\
        s
    \end{pmatrix}$ with the following relation
\begin{equation}\label{eq:transitive}
    \tilde{\bm y}(\bm u) \equiv \bm y(\bm \zeta(\bm u))
\end{equation}
where $\bm y$ is the lensing function defined in Eq \eqref{eq:lenseq}. This transitive lensing function maps our new coordinate system $(t,s)$ to the source plane, shown in Fig. \ref{fig:illus2}. It is convenient to write it as an integral
\begin{align}
    \tilde{\bm y}\left[\begin{pmatrix}
        t\\
        0
    \end{pmatrix}\right] &= \tilde{\bm y}\left[\begin{pmatrix}
        0\\
        0
    \end{pmatrix}\right]  + \int^{t}_{0}dt' \left(\frac{\pd\tilde{\bm y}}{\pd t'}\right)\Bigg|_{(t',0)} \\
    &= \tilde{\bm y}\left[\begin{pmatrix}
        0\\
        0
    \end{pmatrix}\right] + \int^{t}_{0}dt' \left(\frac{\pd \bm y}{\pd \bm x}\right)\Bigg|_{(t',0)}\left(\frac{\pd\bm \zeta}{\pd t'}\right)\Bigg|_{(t',0)} \\
    &= \tilde{\bm y}\left[\begin{pmatrix}
        0\\
        0
    \end{pmatrix}\right] + \int^{t}_{0}dt' \left(\frac{\pd \bm y}{\pd \bm x}\right)\Bigg|_{(t',0)}\left(\frac{d\bm \xi}{d t'}\right).
\end{align}

Now, we can make use of the fact that $\bm \xi$ is a primary curve, which makes $\dfrac{d\bm \xi}{dt'}$ an eigenvector of the Jacobian with the eigenvalue $\lambda_1(t')$. This allows us to write,
\begin{equation}\label{eq:t_integ}
     \tilde{\bm y}\left[\begin{pmatrix}
        t\\
        0
    \end{pmatrix}\right] = \tilde{\bm y}\left[\begin{pmatrix}
        0\\
        0
    \end{pmatrix}\right] + \int^{t}_{0}dt'\,\lambda_1(t')\left(\frac{d\bm \xi}{dt'}\right).
\end{equation}
We can carry out a similar integral along the secondary eigenvector, which is orthogonal to the primary eigenvector. We can write:
\begin{align}
    \tilde{\bm y}\left[\begin{pmatrix}
        t\\
        s
    \end{pmatrix}\right] &= \tilde{\bm y}\left[\begin{pmatrix}
        t\\
        0
    \end{pmatrix}\right]  + \int^{s}_{0}ds' \left(\frac{\pd\tilde{\bm y}}{\pd s'}\right)\Bigg|_{(t,s')} \\
    &= \tilde{\bm y}\left[\begin{pmatrix}
        t\\
        0
    \end{pmatrix}\right] + \int^{s}_{0}ds' \left(\frac{\pd \bm y}{\pd \bm x}\right)\Bigg|_{(t,s')}\left(\frac{\pd\bm \zeta}{\pd s'}\right)\Bigg|_{(t,s')} \\
    &= \tilde{\bm y}\left[\begin{pmatrix}
        t\\
        0
    \end{pmatrix}\right] + \int^{s}_{0}ds' \left(\frac{\pd \bm y}{\pd \bm x}\right)\Bigg|_{(t,s')}
    \begin{pmatrix}
        0 & -1 \\
        1 & 0
    \end{pmatrix}
    \left(\frac{d\bm \xi}{d t}\right)\label{eq:exact_integ}
\end{align}
\newpage
At $s = 0$, $\begin{pmatrix}
        0 & -1 \\
        1 & 0
    \end{pmatrix}
\left(\dfrac{d\bm \xi}{d t}\right)$ is an eigenvector of the Jacobian. However, at $s>0$ this is no longer true in general. We can Taylor expand this integral around $s=0$, keeping terms up to third order in $s$ to write:
\begin{align}
    \tilde{\bm y}\left[\begin{pmatrix}
        t\\
        s
    \end{pmatrix}\right] &= \tilde{\bm y}\left[\begin{pmatrix}
        t\\
        0
    \end{pmatrix}\right] + s \left(\frac{\pd \bm y}{\pd \bm x}\right)\Bigg|_{(t,0)}
    \begin{pmatrix}
        0 & -1 \\
        1 & 0
    \end{pmatrix}
    \left(\frac{d\bm \xi}{d t}\right) \nonumber \\
    &\qquad \qquad + \frac{s^2}{2} \left(\frac{\pd }{\pd s}\frac{\pd \bm y}{\pd \bm x}\right)\Bigg|_{(t,0)}
    \begin{pmatrix}
        0 & -1 \\
        1 & 0
    \end{pmatrix}
    \left(\frac{d\bm \xi}{d t}\right) \nonumber \\
    &\qquad \qquad + \frac{s^3}{6} \left(\frac{\pd^2 }{\pd s^2}\frac{\pd \bm y}{\pd \bm x}\right)\Bigg|_{(t,0)}
    \begin{pmatrix}
        0 & -1 \\
        1 & 0
    \end{pmatrix}
    \left(\frac{d\bm \xi}{d t}\right) \nonumber \\
    &\qquad \qquad+ \mathcal{O}\left[s^4\right].
\end{align}
In Appendix \ref{ap:second_order}, we calculate the partial derivatives of the Jacobian with respect to $s$ in terms of the eigenvalues $\lambda_1$ and $\lambda_2$, along with the orientation angle $\theta$, which allows us to write:
\begin{align}
    \tilde{\bm y}\left[\begin{pmatrix}
        t\\
        s
    \end{pmatrix}\right] 
    &= \tilde{\bm y}\left[\begin{pmatrix}
        t\\
        0
    \end{pmatrix}\right] + s\, \lambda_2(t)
    \begin{pmatrix}
        0 & -1 \\
        1 & 0
    \end{pmatrix}
    \left(\frac{d\bm \xi}{d t}\right) \nonumber \\
    +\frac{s^2}{2}&\left(\lambda_1(t) - \lambda_2(t)\right)\left(\frac{\pd \theta}{\pd s}\right)\Bigg|_{(t,0)} \left(\frac{d \bm \xi}{dt}\right) \nonumber \\ 
    + \frac{s^2}{2}&\left(\frac{\pd \lambda_2}{\pd s}\right)\Bigg|_{(t,0)}
    \begin{pmatrix}
        0 & -1 \\
        1 & 0
    \end{pmatrix}
    \left(\frac{d\bm \xi}{d t}\right) \nonumber \\
    + 
    \frac{s^3}{6}&\left(\lambda_1(t) - \lambda_2(t)\right)\left(\frac{\pd^2 \theta}{\pd s^2}\right)\Bigg|_{(t,0)} \left(\frac{d \bm \xi}{dt}\right)
    \nonumber \\
    +
    \frac{s^3}{6}& 
    \left[\left(\frac{\pd^2 \lambda_2}{\pd s^2}\right)\Bigg|_{(t,0)} + 2(\lambda_1(t) - \lambda_2(t))\left(\frac{\pd \theta}{\pd s}\right)^2\Bigg|_{(t,0)}\right]
    \begin{pmatrix}
        0 & -1 \\
        1 & 0
    \end{pmatrix}
    \left(\frac{d\bm \xi}{d t}\right)
    \nonumber \\
    &+\mathcal{O}\left[s^4\right] \label{eq:s_integ}.
\end{align}
Using Eq. \eqref{eq:xidef}, we can explicitly calculate the derivatives of $\bm \xi$. We have the following set of functions of $t$ that we need to choose to fully parametrize our lens model:
\begin{align}
    1)\quad &\theta_n(t) \text{ is given by Eq. (5)}\\
    2)\quad&\lambda_1(t) \\
    3)\quad&\lambda_2(t) \\
    4)\quad&F(t) \equiv \left(\dfrac{\pd \lambda_2}{\pd s}\right)\Bigg|_{(t,0)}\\
    5)\quad&G(t) \equiv \left(\lambda_1(t) - \lambda_2(t)\right)
    \left(\dfrac{\pd \theta}{\pd s}\right)\Bigg|_{(t,0)} \\
    6)\quad&H(t) \equiv \left(\frac{\pd^2 \lambda_2}{\pd s^2}\right)\Bigg|_{(t,0)} + 2(\lambda_1(t) - \lambda_2(t))\left(\frac{\pd \theta}{\pd s}\right)^2\Bigg|_{(t,0)} \\
    7)\quad&I(t) \equiv \left(\lambda_1(t) - \lambda_2(t)\right)\left(\frac{\pd^2 \theta}{\pd s^2}\right)\Bigg|_{(t,0)}.
\end{align}
Combining Eq. \eqref{eq:s_integ} with Eq. \eqref{eq:t_integ} we can succinctly write the transitive lensing function by using the following notation for the integrals and the derivatives:
\begin{align}
    \mathcal{S}\{f\}(t) &\equiv \int^{t}_0dt' f(t') \\
    \mathcal{D}\{f\}(t) &\equiv \dfrac{df}{dt}.
\end{align}
We arrive at our final expression for the transitive lensing function:
\begin{align}
    \tilde{\bm y}\left[\begin{pmatrix}
        t\\
        s
    \end{pmatrix}\right]
    &=
    \tilde{\bm y}\left[\begin{pmatrix}
        0\\
        0
    \end{pmatrix}\right]
    +
    \begin{pmatrix}
        \cos\theta_0 & -\sin\theta_0 \\
        \sin \theta_0 & \cos\theta_0
    \end{pmatrix}
    \cdot \nonumber \\
    &\left[
    \begin{pmatrix}
        \mathcal{S}\{\lambda_1 \mathcal{D}\{Q_n\}\}(t) \\
        \mathcal{S}\{\lambda_1 \mathcal{D}\{P_n\}\}(t)
    \end{pmatrix}\right.
    +
    s\,\lambda_2(t)
    \begin{pmatrix}
        -\mathcal{D}\{P_n\}(t)\\
        \mathcal{D}\{Q_n\}(t)
    \end{pmatrix}
     \nonumber \\
    &
    +
    \dfrac{s^2}{2}F(t)
    \begin{pmatrix}
        -\mathcal{D}\{P_n\}(t)\\
        \mathcal{D}\{Q_n\}(t)
    \end{pmatrix}
    +
    \frac{s^2}{2}
    G(t)
    \begin{pmatrix}
        \mathcal{D}\{Q_n\}(t) \\
        \mathcal{D}\{P_n\}(t)
    \end{pmatrix}
    \nonumber \\
    & 
    + 
    \frac{s^3}{6}
    H(t)
    \begin{pmatrix}
        -\mathcal{D}\{P_n\}(t)\\
        \mathcal{D}\{Q_n\}(t)
    \end{pmatrix}
    +\left.
    \frac{s^3}{6}
    I(t)
    \begin{pmatrix}
        \mathcal{D}\{Q_n\}(t)\\
        \mathcal{D}\{P_n\}(t)
    \end{pmatrix}
    \right] \nonumber \\
    &+\mathcal{O}\left[s^4\right].\label{eq:y_equation}
\end{align}
However, we must additionally introduce a condition that the curl of the lensing function must vanish: $\nabla \times \bm y = 0$. This is because the angular deflections of a gravitational lens whose extent in the line-of-sight dimension is much smaller than the distance between the observer and the source can be written as the gradient of a lensing potential: $\bm \alpha = \nabla \Psi$. Any line-of-sight effect that will break this condition should be modeled with an extension of our model that includes a curl term. Introducing the vanishing curl condition gives us the following relations which are calculated in Appendix \ref{ap:vanishcurl}:
\begin{align}
    0 &  = G(t) - \mathcal{D}\{\lambda_2\}(t) \\
    0 & = I(t) - \mathcal{D}\{F\}(t) - 3G(t)\mathcal{D}\{\theta\}(t) \\
    0 & = \mathcal{D}\{H\}(t) + 4I(t)\mathcal{D}\{\theta\}(t).
\end{align}
We see that choosing a particular $\lambda_2(t)$ and $F(t)$ also determines  $G(t)$ and $I(t)$, and also $H(t)$ up to a constant $H_0$:
\begin{align}
    G(t) &  =  \mathcal{D}\{\lambda_2\}(t) \\
    I(t) & =   \mathcal{D}\{F\}(t) + 3G(t)\mathcal{D}\{\theta\}(t) \\
    H(t) & = -4\mathcal{S}\{I\mathcal{D}\{\theta\}|(t) + H_0.
\end{align}

\subsection{Conceptual Summary}
Bright lensed arcs typically lie along curves that follow the eigenvectors of the lensing Jacobian (see Fig. \ref{fig:illus3}), which we have named \textit{primary curves}. We first parametrize the primary curve with a polynomial. We then build a lens model that can capture the angular deflections of a complex lens within a neighborhood of the primary curve (see Fig. \ref{fig:illus2}). By parametrizing how the eigenvalues of the lensing Jacobian changes along the primary curve, we parametrize transitive lensing function $\tilde{\bm y}$ (shown in Eq. \eqref{eq:y_equation}). Combining the transitive lensing function with the inverse of the coordinate basis function $\bm \zeta$ (shown in Eq. \eqref{eq:x_equation}) gives us a fully parametric lens model, $\bm y(\bm x) = \tilde{\bm y}\left(\zeta^{-1}(\bm x)\right)$.

\subsection{Degeneracies}\label{sec:degen}
In lensing, one can never directly observe the angular deflections. Instead, the data constists of pixel values of the lensed image of a background source. There can be many combinations of source light distribution and lensing functions that result in the same observed image. Therefore, gravitational lensing is known to host many degeneracies \citep{mass_sheet_degeneracy,Saha_2000,saha_2006, msd_hubble, msd_and_substructure}.

In our lens model, if we make the following transformation $\lambda_1 \rightarrow a \lambda_1$ and $\lambda_2 \rightarrow a \lambda_2$ where $a$ is a constant, we get $\bm{\tilde y} \rightarrow a\bm{\tilde y}$. This means a constant scaling of the source plane, which results in a perfect degeneracy when the source is also scaled with $1/a$, which is known as the \textit{mass-sheet degeneracy} (MST) in the literature \citep{mass_sheet_degeneracy,mst2}. Therefore, we are measuring the relative values of $\lambda_1$ and $\lambda_2$. To obtain the absolute magnifications from any local modeling one needs to add the information from the global mass modeling of the cluster that makes use of all the lensed images of background sources near the cluster.

\subsection{Source Light Modeling}\label{sec:source_light_modeling}

Galaxies can have complex morphologies. Using simple analytical models such as the Sersic profile \citep{sersic} is inadequate in accurately reconstructing the source light distribution. For this reason, it is crucial in lens modeling to have source light models that are flexible enough to capture these complex shapes. 

One common approach is using basis sets such as shapelets \citep{simon_shapelet1} which consists of orthonormal Hermite polynomials. The complexity of a shapelet set can be controlled with the shapelet index $n_\mathrm{max}$ which determines the size of the smallest features relative to the overall size of the source.

In this work, we choose to use an adaptive grid technique very similar to that of \cite{vegetti_delaunay}. Initially, the Hessian matrix $\bm{\mathrm{H}}_i$ of the brightness values at each image pixel $i$ is calculated by finite difference method. The half of the image pixels with the quantity $|\det\bm{\mathrm{H}}|/p_i$ higher than median are chosen to be candidate points, where $p_i$ is the pixel value. These candidate points are subsequently filtered such that the centers of pixels in the final set $\{\bm q_i\}$ are further from each other than 2 pixel widths. These center points are then fed into the lens model to form a set of points on the source plane: $\{\bm w_i\}$ with $\bm w_i = \bm y (\bm q_i)$. This set of points make up the vertices of a Delaunay triangulation on the source plane. Each vertex $\bm w_i$ is assigned an amplitude $c_i$ which determines the surface brightness of the source at that position. For the surface brightness at a generic point on the source plane $\bm w$, we first calculate which Delaunay triangle $\bm w$ is enclosed within. Then the surface brightness $I_s(\bm w)$ is given by a linear combination of the amplitudes $(c_i,c_j,c_k)$ of the three vertices $(\bm w_i, \bm w_j, \bm w_k)$ of that triangle:
\begin{align}
    I_s(\bm w) = &c_i D_{ijk}(\bm w) \\
    + &c_j D_{kij}(\bm w) \\ 
    + &c_k D_{jki}(\bm w)
\end{align}
where
\begin{align}
     D_{ijk}(\bm w) &\equiv \frac{(\bm w_1 - \bm w_{k1})(\bm w_{j2} - \bm w_{k2}) - (\bm w_{j1} - \bm w_{k1})(\bm w_2 - \bm w_{k2})}{(\bm w_{i1} - \bm w_{k1})(\bm w_{j2} - \bm w_{k2}) - (\bm w_{j1} - \bm w_{k1})(\bm w_{i2} - \bm w_{k2})}.
\end{align}
If a point $\bm w$ falls outside of any Delaunay triangle, it is assigned the surface brightness of the nearest vertex. 

A set of source vertex amplitudes $\bm c \equiv (c_1,c_2,...,c_{n_s})$ fully determines source surface brightness $S_\mathrm{source}(\bm c)$ on the source plane. Subsequently, a lensing operation $\mathcal{L}(\bm m)$ controlled by a set of lens model parameters represented as the vector $\bm m$ fully determines the surface brightness $\bm I$ on the image plane with
\begin{align}
    \bm I(\bm m;\bm c) = \mathcal{L}(\bm m)\{S(\bm c)\}
\end{align}
To convert this surface brightness to an observed image reconstruction, we need convolve it with a point spread function (PSF) then subsequently sample it by a pixellated grid. We denote these operations with $\mathcal{T}$. So our model reconstruction can be represented as a vector
\begin{align}
    \bm M(\bm m;\bm c) = \mathcal{T}\{\mathcal{L}(\bm m)\{S(\bm c)\}\}.
\end{align}
An important property of the model reconstruction is that it is a linear function of $\bm c$, i.e. $\pd^2 \bm M/\pd c_i \pd c_j = 0$. For a given $\bm m$ there is a matrix $\bm A(\bm m)$ that gives $\bm M(\bm m;\bm c) = \bm A(\bm m) \bm c$. Our data which containts the image pixels of the lensed arc that we are modeling can also be represented as a vector: $\bm d$. Solving for $\bm c$ is then the simple problem of minimizing the following loss function
\begin{align}
    \chi^2(\bm c) = [\bm d - \bm A(\bm m)\bm c]^T \Sigma^{-1} [\bm d - \bm A(\bm m)\bm c]
\end{align}
where $\Sigma$ is the covariance matrix of the data vector, assuming that the pixel errors are Gaussian. We remind that $\bm c$ represents the surface brightness at the vertices of the Delaunay triangulation. A negative value of any $c_i$ would mean we have an unphysical source model as surface brightness cannot be negative. So given our data and a set of lens model parameters, the source amplitudes are given by
\begin{align}
    \text{arg min} \quad \chi^2(\bm c) \quad \text{subject to} \quad \bm c_i \geq 0 \quad \forall i \label{eq:c_optim}
\end{align}
This is a well known problem called \textit{quadratic optimization}, and it has many efficient solvers. We use the function \texttt{scipy.optimize.nnls} which is based on the method developed by \cite{nonneg}.

\begin{figure*}
    \centering
    \includegraphics[width=0.75\textwidth]{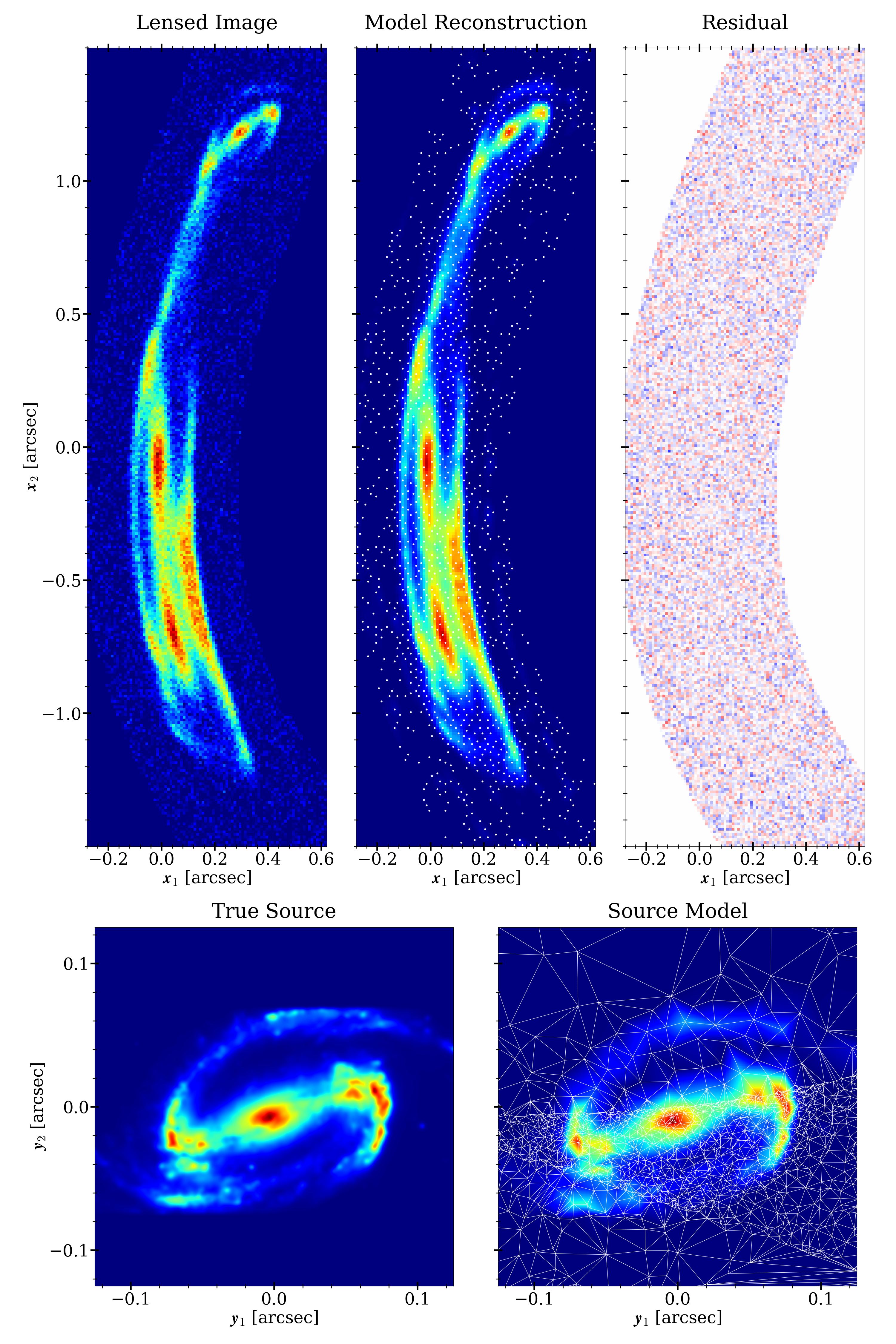}
    \caption{Source Reconstruction of a Mock Lens. \textit{Top Left: } 
    The lensed image.
    \textit{Top middle:}
    Model reconstruction with the white points showing the source modeling points. These white dots are the points that are mapped onto the source plane to form the Delaunay triangulation vertices as described in \S \ref{sec:source_light_modeling}.
    \textit{Top right:}
    Residuals after the model reconstruction is subtracted from the image.
    \textit{Bottom left:}
    The true source that is used in creating the mock data image.
    \textit{Bottom right:}
    The source model reconstruction with the white lines showing the Delaunay triangulation. The vertices of the triangulation are the positions after the points in the top middle subplot are mapped onto the source plane using the lens model.}
    \label{fig:source_model}
\end{figure*}

We illustrate this pipeline for source reconstruction in Fig. \ref{fig:source_model}. We create a simulated lensed image using the \textit{Hubble Space Telescope} (HST) image of galaxy NGC 1300, which is shown in the \textit{bottom left} subplot. The simulated lensed data is shown in the \textit{top left} subplot. Following the procedure described earlier in this section, we select a set of points $\{\bm q_i\}$ on the image plane, which are shown in the \textit{top middle} subplot as the white dots. These points are then projected onto the source plane to form the vertices of a Delaunay triangulation, which is shown in the \textit{bottom right} subplot as the white triangles. The vertex amplitudes $\bm c$ are calculated using Eq. \eqref{eq:c_optim}. The \textit{bottom right} and the \textit{top middle} subplots show the source surface brightness reconstruction and the image model reconstruction respectively. Comparing the source reconstruction to the true source, we see that we are able to recover many of the features seen in the complex morphology of NGC 1300. In the \textit{top right} subplot, we show the residuals after subtracting the image reconstruction from the mock data.

\subsection{Nested Sampling and Bayesian Inference}

The dependence of the model reconstruction $\bm M(\bm m;\bm c)$ on the lens model parameters $\bm m$ is highly non-linear. For each $\bm m$, we set $\bm c$ to be the solution of Eq. \eqref{eq:c_optim}. We can write this as $\bm M(\bm m) \equiv M(\bm m; \bm c_\mathrm{min})$. Assuming that the pixel errors are Gaussian, we obtain the posteriors of the model parameters using Bayesian inference:
\begin{align}
    P(\bm d|\bm m) = \frac{\exp\left[-\frac{1}{2}\left(\bm d - \bm M(\bm m)\right)^T \Sigma^{-1}\left(\bm d - \bm M(\bm m)\right)\right]}{\sqrt{(2\pi)^{\dim \bm d}\det(\Sigma)}}
\end{align}
\begin{align}
        P(\bm m| \bm d) = \frac{P(\bm d| \bm m)P(\bm m)}{P(\bm d)}.
\end{align}
We use the nested sampling package \texttt{dynesty} \citep{DYNESTY} to sample the likelihood function to obtain the lens parameter posteriors, along with the best-fit parameters.
 
\section{Results}\label{sec:results}
\subsection{SDSS J1110+6459}\label{sec:arc1}
SDSS J1110+6459 is a galaxy cluster gravitational lens that is host to one of the most striking examples of a lensed arc, which can be seen in Fig. 3 of \cite{SDSSJ1110+6459_star_formation}. This arc consists of three merging images of a background galaxy at z=2.481 \citep{2013MNRAS.436.1040S,SDSSJ1110+6459_star_formation}. The galaxy cluster lens is at redshift z=659 \citep{2012MNRAS.420.3213O,2013MNRAS.436.1040S,SDSSJ1110+6459_star_formation}.

An early strong lensing analysis of SDSS J1110+6459 is \cite{2012MNRAS.420.3213O}, where they used the low resolution ground-based \textit{Subaru} telescope. Without the accurate spectroscopic redshifts, these early lens modeling efforts are poorly constrained. Using the higher-resolution HST imaging, as well as the spectroscopic redshifts for the sources, \cite{SDSSJ1110+6459_star_formation} provided the first high-fidelity strong lens modeling, which we use when we make our parametrization choices for our lens model.

\begin{figure*}
    \centering
    \includegraphics[width=0.80\textwidth]{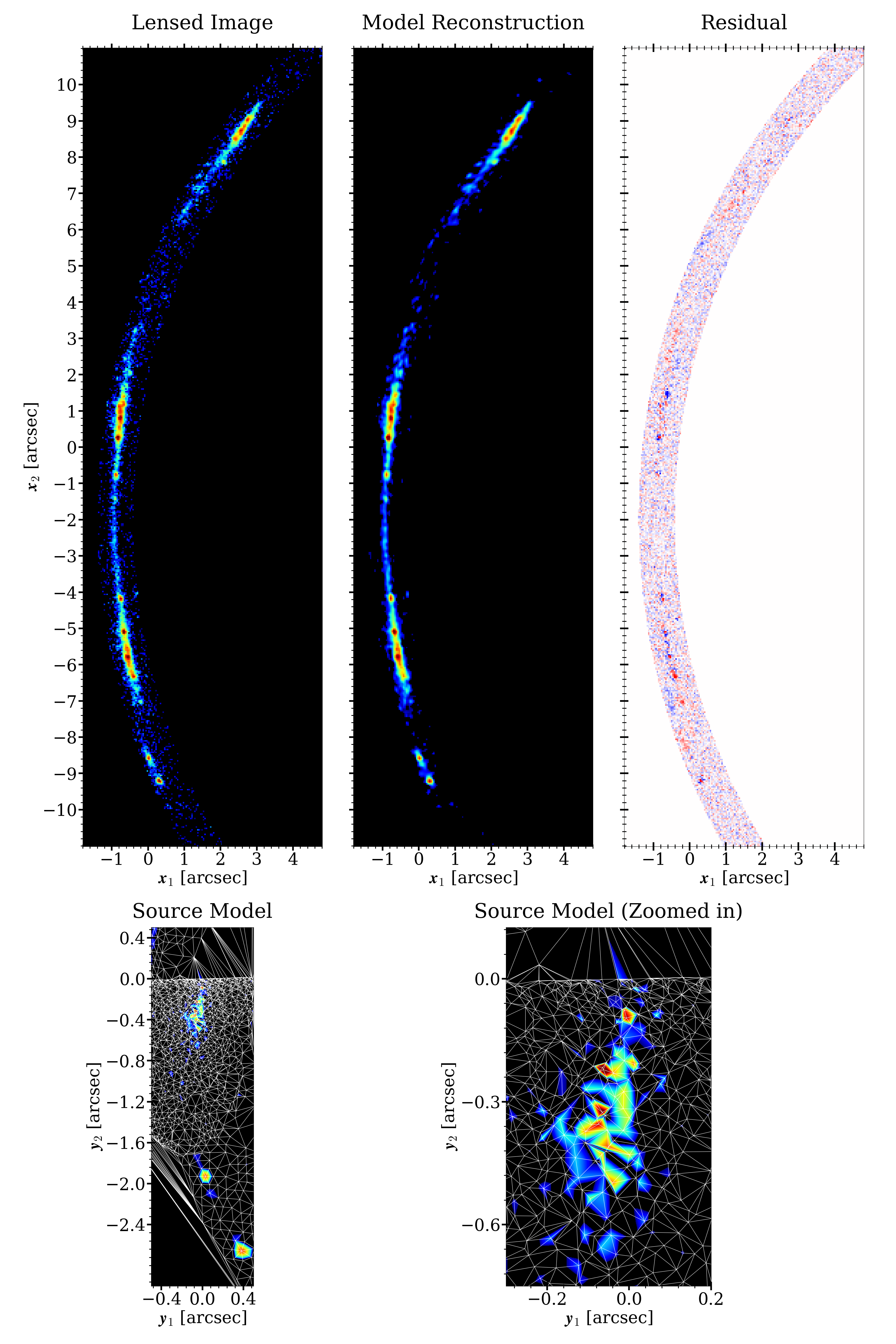}
    \caption{Image Reconstruction of the bright arc in SDSS J1110+6459. \textit{Top Left: } 
    The lensed image cutout of the bright arc from the HST/WFC3 UVIS F390W observation.
    \textit{Top middle:}
    Model reconstruction.
    \textit{Top right:}
    Residuals after the model reconstruction is subtracted from the data
    \textit{Bottom left:}
    The source model reconstruction with the white lines showing the Delaunay triangulation. The vertices of the triangulation are the positions after the points in the top middle subplot are mapped onto the source plane using the lens model.
    \textit{Bottom right:}
    The source model reconstruction zoomed in showing high magnification regions.}
    \label{fig:sdssj1110model}
\end{figure*}

We show the cutout we use of the bright arc in SDSS J1110+6459 in the \textit{top left} subplot of Fig. \ref{fig:sdssj1110model}. The data is taken with HST/WFC3 on January 8th 2013 and is publicly available on the MAST portal\footnote{\label{note1}\url{https://mast.stsci.edu/portal/Mashup/Clients/Mast/Portal.html}}. We use the F390W filter observation with 1212 seconds of exposure. The three images of the same source galaxy are clearly seen from the bottom to the top along our cutout. Something very qualitatively similar to what we have shown in Fig. \ref{fig:illus3} is happening in this arc. The primary curve lies roughly along our cutout with the primary eigenvalue $\lambda_1$ flipping signs twice as we move from the bottom of the image to the top. The first sign flip (from positive to negative) is between the first image and the second image near $(-0.5'',-2.5'')$ which we use as initialization of $(d_1,d_2)$ (shown in Eq. \ref{eq:xidef}). The second sign flip is (from negative to positive) between the second image and the third image near $(1.0'',5.0'')$.

Considering these properties, we parametrize $\lambda_1$ as a polynomial
\begin{align}
    \lambda_1(t) &= r_0 t \left(1- \frac{t}{t_\mathrm{cr}}\right)\left(1- \frac{t}{t_\mathrm{cr}}-\frac{r_\mathrm{cr}}{r_0}\frac{t}{t_\mathrm{cr}}\right) \nonumber \\
    &\quad + q_4 t^2\left(1 - \frac{t}{t_\mathrm{cr}}\right)^2\left(1 + q_5 t + q_6 t^2 + ... + q_m t^{m-4}\right) \label{eq:lambda1model}
\end{align}
which has the following properties:
\begin{align}
    1)&\quad \lambda_1(0) = 0, \label{eq:prop1}\\
    2)&\quad \lambda_1(t_\mathrm{cr}) = 0,\label{eq:prop2} \\
    3)&\quad \frac{d\lambda_1}{dt}\Bigg|_{t = 0} = r_0, \label{eq:prop3}\\
    4)&\quad \frac{d\lambda_1}{dt}\Bigg|_{t = t_\mathrm{cr}} = r_\mathrm{cr}. \label{eq:prop4}
\end{align}
The parameter $r_0$ is the slope of $\lambda_1$ at $t=0$, which has a negative value as the first sign flip is from positive to negative. The parameter $t_\mathrm{cr}$ is the position of the second sign flip of $\lambda$, and $r_\mathrm{cr}$ is the slope of $\lambda_1$ at $t= t_\mathrm{cr}$, which has a positive value as the second sign flip is from negative to positive. The remaining parameters $q_4,...,q_m$ are higher-order terms that allow further flexibility in $\lambda_1$ while preserving the properties in Eq.(\ref{eq:prop1}-\ref{eq:prop4}). We fix $\lambda_2(0) = 1$ due to the degeneracy described in \S \ref{sec:degen}, which means that the values that we infer in our analysis for $\lambda_1(t)$ are ratios of $\lambda_1(t)/\lambda_2(0)$.

Our lens model has the parameters for the primary curve $(d_1,d_2,\theta_0,a_1,...,a_n)$ and the parameters for the eigenvalue $(t_\mathrm{cr},r_0,r_\mathrm{cr},q_4,...,q_m)$. We chose fiducial values of $n = 3$ and $m=6$ for our analysis of SDSS J1110+6459, as this analysis is to show the fidelity of our method. In a more detailed analysis, one should optimize $n$ and $m$ using Bayesian Information Criterion (BIC), which we left for future work as it is computationally costly.

We set uninformative flat priors for all the model parameters. The priors for $(d_1,d_2)$ are chosen such that it stays between the first and the second image along the arc, as the sign flip for $\lambda_1$ needs to be between these images. Similarly, the prior $t_\mathrm{cr}$ is chosen such that the second sign flip is between the second and third images along the arc. The priors for $r_0$ and $r_\mathrm{cr}$ are chosen such that the former is strictly negative and the latter is strictly positive. We calculate the posterior probabilities for the model parameters using the nested sampling package \texttt{dynesty} \citep{DYNESTY}.

In Fig. \ref{fig:sdssj1110model}, we show the best fit of our lens model obtained from our sampling. In the \textit{top left} subplot, we show the lensed image of the bright arc. In the \textit{top middle} subplot we show our model reconstruction. In the \text{top right} subplot, we show the image residuals after the reconstruction is subtracted from the data. In the \text{bottom left} subplot, we show the source reconstruction with the Delaunay triangulation shown as white lines. In the \text{bottom left} subplot we show a zoomed-in version of our source reconstruction focusing on a high magnification region.

\subsection{SDSS J0004-0103 }

SDSS J0004-0103 is another galaxy cluster gravitational lens that also hosts a clear lensed arc that consists of 3 merging images of a background galaxy at z=1.681 \citep{rigby2018}.
An earlier lensing analysis of SDSS J0004-0103 \citep{sharon_2020} resulted in a non-unique lens model with large systematic uncertainty due to the limited number of background sources and a lack of a dominant cluster center galaxy.

\begin{figure*}
    \centering
    \includegraphics[width=0.89\textwidth]{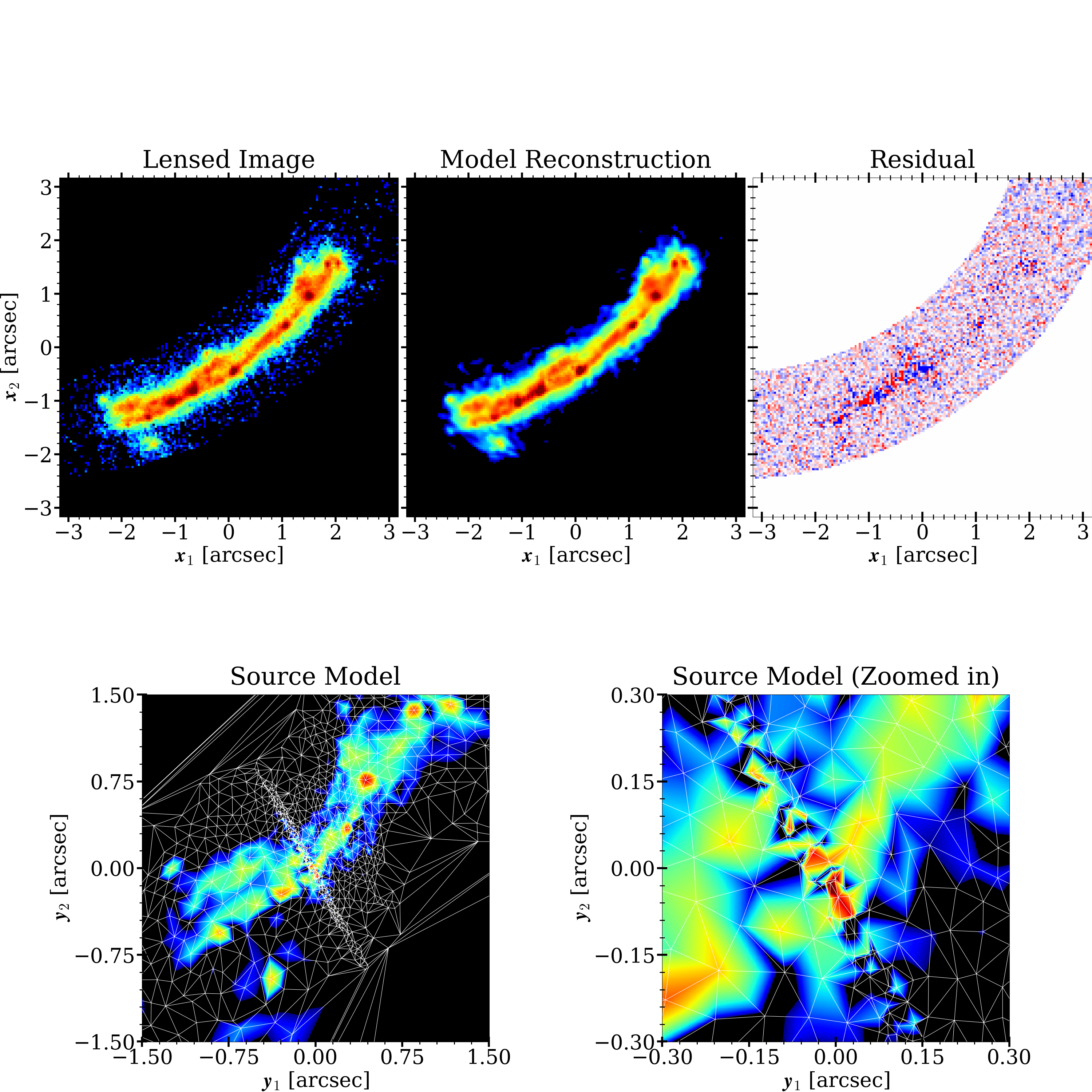}
    \caption{Image Reconstruction of the bright arc in SDSS J0004-0103. \textit{Top Left: } 
    The lensed image cutout of the bright arc from the HST/WFC3 UVIS F390W observation.
    \textit{Top middle:}
    Model reconstruction.
    \textit{Top right:}
    Residuals after the model reconstruction is subtracted from the data
    \textit{Bottom left:}
    The source model reconstruction with the white lines showing the Delaunay triangulation. The vertices of the triangulation are the positions after the points in the top middle subplot are mapped onto the source plane using the lens model.
    \textit{Bottom right:}
    The source model reconstruction zoomed in showing high magnification regions.}
    \label{fig:sdssj1004model}
\end{figure*}

We show the cutout we use of the bright arc in SDSS J1110+6459 in the \textit{top left} subplot of Fig. \ref{fig:sdssj1004model}. The data was taken with HST/WFC3 on September 24th, 2013 and is publicly available on the MAST portal$^{\ref{note1}}$. We use the F390W filter observation with 2388 seconds of exposure. \cite{sharon_2020} has shown that the arc crosses the critical curve twice, resulting in two sign flips of $\lambda_1$. The blend of the three images is more difficult to see by eye in this system compared to the arc described in \S \ref{sec:arc1}. Therefore, we use the mathematical model for $\lambda_1$ shown in Eq. \eqref{eq:lambda1model}. Similar to \S \ref{sec:arc1}, we chose fiducial values of $n = 3$ and $m=6$ for our analysis of SDSS J0004-0103, as this analysis is to show the fidelity of our method. We leave the full BIC optimization as future work.

In Fig. \ref{fig:sdssj1004model} we show the best fit of our lens model obtained from our sampling. In the \textit{top left} subplot, we show the lensed image of the bright arc. In the \textit{top middle} subplot we show our model reconstruction. In the \text{top right} subplot, we show the image residuals after the reconstruction is subtracted from the data. In the \text{bottom left} subplot, we show the source reconstruction with the Delaunay triangulation shown as white lines. In the \text{bottom left} subplot, we show a zoomed-in version of our source reconstruction focusing on a high magnification region.

\section{Discussion}\label{sec:discussion}

Bright lensed arcs provide a rare opportunity to probe structure down to 100 pc scales for high redshift galaxies. Probing these small scales gives us vital information on the formation and evolution of galaxies. Having an accurate lens model near the critical curves where the magnification is highest is crucial in properly de-lensing the image to infer what the source looks like. The simultaneous source and lens modeling we have developed in this work allows for better source reconstructions of these bright arcs.

The global cluster lens models are shown to benefit drastically by utilizing the information of the extended source light distributions \citep{Pascale_2022,bergamini_2021,diego_2022,sharon_2022}. Using the constraints from the local lens model that we developed in this work for the bright arcs is similarly likely to improve the fidelity of global lens models as it utilizes the full information in the extended light distribution.

The lens model that we have used throughout this work is smooth by construction. It has the flexibility to capture complicated lensing functions, but it can only fit the slowly varying angular deflection field of the galaxy cluster. In the Cold Dark Matter (CDM) paradigm, galaxy halos and cluster halos are expected to host a large number of low-mass subhalos (<$10^{10} \mathrm{M_\odot}$) that do not have detectable luminosities. If properly aligned, these subhalos, along with similarly low mass dark line-of-sight halos, create small perturbations in bright arcs with their angular deflections. In galaxy-galaxy lenses, lens modeling by fitting the pixels of the image has been shown to have the sensitivity to detect these dark perturbers down to $10^{8} \mathrm{M_\odot}$ \citep{vegetti2010,vegetti2012}. Recently, it has been shown that the same sensitivity can be reached in cluster lenses by locally modeling the angular deflections near the multiple images of a source \citep{sengul_cab}. By modeling the smooth component of the angular deflections of the cluster with the method that we developed in this work, one can similarly detect low-mass perturbers in cases where there are significant residuals demanding a localized mass to be added to the model. We leave this hunt for dark substructure for future work.

\begin{acknowledgements}
      AÇŞ is supported by the Samuel P. Langley PITT PACC Postdoctoral Fellowship. AÇŞ would like to thank Simon Birrer, Cora Dvorkin, Chandrika Chandrashekar, Nino Ephremidze, and Andrew Zentner for useful discussions and comments.
\end{acknowledgements}

\bibliographystyle{mnras}
\bibliography{main}

\appendix

\section{Quick Inversion}\label{sec:quick_inv}
It is useful to also have a way to quickly invert $\zeta$ in Eq. \eqref{eq:x_equation} to compuationally implement lensing calculations. We start with a point
\begin{equation}
    \begin{pmatrix}
        x_1\\
        x_2
    \end{pmatrix}
    =
    \hat x_1
    \begin{pmatrix}
        \cos\theta_0\\
        \sin\theta_0
    \end{pmatrix}
    +
    \hat x_2
    \begin{pmatrix}
        -\sin\theta_0\\
        \cos\theta_0
    \end{pmatrix}.
\end{equation}
For a given $(x_1,x_2)$ we can easily find $(\hat x_1, \hat x_2)$. It is simply,
\begin{align}
    \hat x_1 &= x_1\cos\theta_0 + x_2\sin\theta_0\\
    \hat x_2 &= x_2\cos\theta_0 - x_1\sin\theta_0.
\end{align}
We try to solve the inverse function $\bm \zeta^{-1}$ that gives,
\begin{equation}
    \bm \zeta^{-1} \left[\begin{pmatrix}
        x_1\\
        x_2
    \end{pmatrix}\right] =
    \begin{pmatrix}
        t\\
        s
    \end{pmatrix}.
\end{equation}
We can first set $(t,0)$ and find the $t$ that gives the nearest point. To do this we write,
\begin{align}
    f(t) &= \left|-\begin{pmatrix}
        x_1\\
        x_2
    \end{pmatrix} + 
    Q_n(t)
    \begin{pmatrix}
        \cos\theta_0\\
        \sin\theta_0
    \end{pmatrix}
    +
    P_n(t)
        \begin{pmatrix}
        -\sin\theta_0 \\
        \cos\theta_0
    \end{pmatrix}
    \right|^2 \\
    &= 
    \left|\begin{pmatrix}
        (Q_n(t)-\hat x_1)\cos\theta_0 - (P_n(t)-\hat x_2) \sin\theta_0\\
        (Q_n(t)-\hat x_1)\sin\theta_0 + (P_n(t)- \hat x_2) \cos\theta_0
    \end{pmatrix}
    \right|^2 \\
    &=  (Q_n(t)-\hat x_1)^2 + (P_n(t)-\hat x_2)^2.
\end{align}
Candidate solutions for the nearest point are the roots of the derivative of $f(t)$. We calculate,
\begin{equation}
    D\{f\}(t) = 2(Q_n(t)-\hat x_1)\mathcal{D}\{Q_n\}(t) + 2(P_n(t)-\hat x_2)\mathcal{D}\{P_n\}(t).
\end{equation}
Among the roots of $D\{f\}$, we select the $t_\mathrm{min}$ that gives the minimum $f(t_\mathrm{min})$. This root finding is quickly done with Newton's method by iterating:
\begin{align}
    t_0 &= 0 \\
    t_{i+1} &= t_{i} - \frac{\mathcal{D}\{f\}(t_i)}{\mathcal{DD}\{f\}(t_i)}.
\end{align}
We find that $i=10$ gives the desired accuracy of $|t_\mathrm{min} - t_{i}| < 10^{-8}$. Once this root is found, the value of $s$ is simply given by
\begin{equation}
    s = \frac{\hat{x}_2 - P_n(t_\mathrm{min})}{\mathcal{D}\{Q_n\}(t_\mathrm{min})} = -\frac{\hat{x}_1 - Q_n(t_\mathrm{min})}{\mathcal{D}\{P_n\}(t_\mathrm{min})},
\end{equation}
whichever is not singular.

\section{Secondary Eigenvector Approximation}\label{ap:second_order}
To calculate the transitive lensing function for larger $s$, we start by writing all the terms of Eq. \eqref{eq:exact_integ} up to $s^3$:
\begin{align}
    \tilde{\bm y}\left[\begin{pmatrix}
        t\\
        s
    \end{pmatrix}\right]
    &= \tilde{\bm y}\left[\begin{pmatrix}
        0\\
        0
    \end{pmatrix}\right] + \int^{t}_{0}dt'\,\lambda_1(t')\left(\frac{d\bm \xi}{dt'}\right) \\
    &\qquad \qquad + \int^{s}_{0}ds' \left(\frac{\pd \bm y}{\pd \bm x}\right)
    \begin{pmatrix}
        0 & -1 \\
        1 & 0
    \end{pmatrix}
    \left(\frac{d\bm \xi}{d t}\right) \nonumber \\
    &= \tilde{\bm y}\left[\begin{pmatrix}
        0\\
        0
    \end{pmatrix}\right] + \int^{t}_{0}dt'\,\lambda_1(t')\left(\frac{d\bm \xi}{dt'}\right) \nonumber \\
    &\qquad \qquad+ s \left(\frac{\pd \bm y}{\pd \bm x}\right)\Bigg|_{(t,0)}
    \begin{pmatrix}
        0 & -1 \\
        1 & 0
    \end{pmatrix}
    \left(\frac{d\bm \xi}{d t}\right) \nonumber \\
    &\qquad \qquad + \frac{s^2}{2} \left(\frac{\pd }{\pd s}\frac{\pd \bm y}{\pd \bm x}\right)\Bigg|_{(t,0)}
    \begin{pmatrix}
        0 & -1 \\
        1 & 0
    \end{pmatrix}
    \left(\frac{d\bm \xi}{d t}\right) \nonumber \\
    &\qquad \qquad + \frac{s^3}{6}\left(\frac{\pd^2 }{\pd s^2} \frac{\pd \bm y}{\pd \bm x}\right)\Bigg|_{(t,0)}
    \begin{pmatrix}
        0 & -1 \\
        1 & 0
    \end{pmatrix}
    \left(\frac{d\bm \xi}{d t}\right) 
    + \mathcal{O}\left[s^4\right].
\end{align}
Using Eq. \eqref{eq:jacob2} we calculate the partial derivative:
\begin{align}
    \left(\frac{\pd}{\pd s}\frac{\pd \bm{y}}{\pd \bm{x}}\right) = 
    &\frac{\pd\theta}{\pd s}
    \begin{pmatrix}
        -\sin \theta & - \cos \theta \\
        \cos \theta & -\sin \theta 
    \end{pmatrix}
    \begin{pmatrix}
        \lambda_1 & \\
        & \lambda_2 
    \end{pmatrix}
    \begin{pmatrix}
        \cos \theta & \sin \theta \\
        -\sin \theta & \cos \theta 
    \end{pmatrix} \nonumber \\
    &+ 
    \begin{pmatrix}
        \cos \theta & - \sin \theta \\
        \sin \theta & \cos \theta 
    \end{pmatrix}
    \begin{pmatrix}
        \dfrac{\pd \lambda_1}{\pd s} & \\
        & \dfrac{\pd \lambda_2}{\pd s}
    \end{pmatrix}
    \begin{pmatrix}
        \cos \theta & \sin \theta \\
        -\sin \theta & \cos \theta 
    \end{pmatrix} \nonumber \\
    &+ 
    \frac{\pd\theta}{\pd s}
    \begin{pmatrix}
        \cos \theta & - \sin \theta \\
        \sin \theta & \cos \theta 
    \end{pmatrix}
    \begin{pmatrix}
        \lambda_1 & \\
        & \lambda_2 
    \end{pmatrix}
    \begin{pmatrix}
        -\sin \theta & \cos \theta \\
        -\cos \theta & -\sin \theta 
    \end{pmatrix}.
\end{align}
As this matrix appears in the calculation as being applied to $  \begin{pmatrix}
    0 & -1 \\
    1 & 0
\end{pmatrix}
\begin{pmatrix}
    \cos \theta \\
    \sin \theta
\end{pmatrix}$, we calculate:
\begin{align}
  \left(\frac{\pd}{\pd s}\frac{\pd \bm{y}}{\pd \bm{x}}\right)
    \begin{pmatrix}
        -\sin \theta \\
        \cos \theta
    \end{pmatrix}
    &=
    \frac{\pd\theta}{\pd s}
    \begin{pmatrix}
        -\sin \theta & - \cos \theta \\
        \cos \theta & -\sin \theta 
    \end{pmatrix}
    \begin{pmatrix}
        \lambda_1 & \\
        & \lambda_2 
    \end{pmatrix}
    \begin{pmatrix}
        0 \\
        1
    \end{pmatrix}
    \nonumber \\
    &\quad + 
    \begin{pmatrix}
        \cos \theta & - \sin \theta \\
        \sin \theta & \cos \theta 
    \end{pmatrix}
    \begin{pmatrix}
        \dfrac{\pd \lambda_1}{\pd s} & \\
        & \dfrac{\pd \lambda_2}{\pd s}
    \end{pmatrix}
    \begin{pmatrix}
        0 \\
        1
    \end{pmatrix}
    \nonumber \\
    &\quad + 
    \frac{\pd\theta}{\pd s}
    \begin{pmatrix}
        \cos \theta & - \sin \theta \\
        \sin \theta & \cos \theta 
    \end{pmatrix}
    \begin{pmatrix}
        \lambda_1 & \\
        & \lambda_2 
    \end{pmatrix}
    \begin{pmatrix}
        1 \\
        0
    \end{pmatrix} \\
    &=
    \frac{\pd\theta}{\pd s}
    \begin{pmatrix}
        -\sin \theta & - \cos \theta \\
        \cos \theta & -\sin \theta 
    \end{pmatrix}
    \begin{pmatrix}
        0 \\
        \lambda_2
    \end{pmatrix}
    \nonumber \\
    &\quad + 
    \begin{pmatrix}
        \cos \theta & - \sin \theta \\
        \sin \theta & \cos \theta 
    \end{pmatrix}
    \begin{pmatrix}
        0 \\
        \\
        \dfrac{\pd \lambda_2}{\pd s}
    \end{pmatrix}
    \nonumber \\
    &\quad + 
    \frac{\pd\theta}{\pd s}
    \begin{pmatrix}
        \cos \theta & - \sin \theta \\
        \sin \theta & \cos \theta 
    \end{pmatrix}
    \begin{pmatrix}
        \lambda_1\\
        0
    \end{pmatrix} \\
    &=
    \left(\lambda_1 - \lambda_2  \right)\frac{\pd\theta}{\pd s}
    \begin{pmatrix}
        \cos \theta \\
        \sin \theta 
    \end{pmatrix}
    + 
    \dfrac{\pd \lambda_2}{\pd s}
    \begin{pmatrix}
        - \sin \theta \\
        \cos \theta 
    \end{pmatrix}.
\end{align}
We can do the same calculation for the second derivative which gives us
\begin{align}
    \left(\frac{\pd^2}{\pd s^2} \frac{\pd \bm y}{\pd \bm x}\right)
    \begin{pmatrix}
        -\sin\theta\\
        \cos\theta
    \end{pmatrix} &=\left[\left(\frac{\pd ^2 \theta}{\pd s^2}\right)\left(\lambda_1 - \lambda_2\right)\right]
    \begin{pmatrix}
        \cos\theta\\
        \sin\theta
    \end{pmatrix}\\
    &\quad
    + \left[ 2\left(\frac{\pd \theta}{\pd s}\right)^2 \left(\lambda_1 - \lambda_2\right)
    + \frac{\pd^2 \lambda_2}{\pd s^2}\right]
        \begin{pmatrix}
        -\sin \theta \\
        \cos \theta
    \end{pmatrix}.
\end{align}
\section{Enforcing a Vanishing Curl}\label{ap:vanishcurl}
To calculate the conditions for $\bm y$ to have a vanishing curl, we start by writing the Jacobian of the lensing function. We use the chain rule for multivariate functions:
\begin{align}
    \frac{\pd \bm y}{\pd \bm x} &= \left[\frac{\pd \tilde{\bm y}}{\pd \bm u}\right]\left[\frac{\pd \bm u}{\pd \bm x}\right] =  \left[\frac{\pd \tilde{\bm y}}{\pd \bm u}\right]\left[\frac{\pd \bm x}{\pd \bm u}\right]^{-1} \\
    &= \dfrac{1}{\dfrac{\pd x_1}{\pd t} \dfrac{\pd x_2}{\pd s} - \dfrac{\pd x_1}{\pd s} \dfrac{\pd x_2}{\pd t}}
    \begin{pmatrix}
        \dfrac{\pd \tilde y_1}{\pd t} & \dfrac{\pd \tilde y_1}{\pd s}\\
        &\\
        \dfrac{\pd \tilde y_2}{\pd t} & \dfrac{\pd \tilde y_2}{\pd s}
    \end{pmatrix}
    \begin{pmatrix}
        \dfrac{\pd x_2}{\pd s} & -\dfrac{\pd x_1}{\pd s}\\
        &\\
        -\dfrac{\pd x_2}{\pd t} & \dfrac{\pd x_1}{\pd t}
    \end{pmatrix}.
\end{align}
We see that the curl is proportional to
\begin{align}
    \nabla \times \bm y &\propto \frac{\pd \tilde{\bm{ y}}}{\pd s}\cdot \frac{\pd \bm x}{\pd t} - \frac{\pd \tilde{\bm{ y}}}{\pd  t}\cdot \frac{\pd \bm x}{\pd s}.
\end{align} 
Using Eq. \eqref{eq:x_equation} and \eqref{eq:y_equation} to calculate the partial derivatives, we see that a vanishing curl implies:
\begin{align}
    \nabla \times \bm y = 0
    \quad &\Rightarrow\quad 0 = G(t) - \mathcal{D}\{\lambda_2\}(t) \label{eq:curlfree_cond1}
\end{align}
and
\begin{align}
    \nabla \times \bm y = 0 \quad \Rightarrow \quad  0 &= - \mathcal{D}\left\{F\right\}(t) -3G(t)
    \mathcal{D}\{\theta_n\}(t) \label{eq:curlfree_cond2}.
\end{align}
We see that we are not free to choose $G(t)$ or $F(t)$. They are constrained by our choice of $\lambda_2(t)$. Combining Eq. \eqref{eq:curlfree_cond1} and \eqref{eq:curlfree_cond2} gives us the following solutions for $G$ and $F$:
\begin{align}
    G(t) &= \mathcal{D}\{\lambda_2\}(t) \\
    F(t) &= -3\mathcal{S}\{\mathcal{D}\{\lambda_2\}\mathcal{D}\{\theta_n\}\}(t) + F_0
\end{align}
where $F_0$ is a constant that represents the second partial derivative of $\lambda_2$ at $t=0$ with respect to $s$.

\section{Properties of the Lens Model}

\subsection{Magnification}
We calculate the inverse magnification of our model as a function of $(t,s)$:
\begin{align}
    \frac{1}{\mu} &= \det \left[\frac{\pd \bm y}{\pd \bm x}\right] = \frac{\det\left[\dfrac{\pd \bm y}{\pd \bm u}\right]}{\det\left[\dfrac{\pd \bm x}{\pd \bm u}\right]} \\
    &=\frac{\dfrac{\pd y_1}{\pd t}\dfrac{\pd y_2}{\pd s} - \dfrac{\pd y_1}{\pd s} \dfrac{\pd y_2}{\pd t}}{\dfrac{\pd x_1}{\pd t}\dfrac{\pd x_2}{\pd s} - \dfrac{\pd x_1}{\pd s}\dfrac{\pd x_2}{\pd t}}.
\end{align}
For the simple case of when $\lambda_2 = \text{const}$, $F = 0$, and $H_0 = 0$, we get:
\begin{align}
    \frac{1}{\mu} = \frac{\lambda_1(t) \lambda_2 - s\lambda_2^2 \mathcal{D}\{\theta_n\}(t)}{1 - s \mathcal{D}\{\theta_n\}(t)}.
\end{align}
When $s=0$ we see that the inverse magnification is simply $\mu^{-1} = \lambda_1 \lambda_2$ as expected.

\section{Various Limits of the Lens Model}
\subsection{Linear Lensing}
In the simple case where $\theta_n(t) = 0$, $\lambda_1(t) = \text{const}$, $\lambda_2(t) = \text{const}$, $F(t) = 0$, and $H_0 = 0$ the lens model reduces to a simple shear and convergence given by,
\begin{align}
    \frac{\pd \bm{y}}{\pd \bm{x}} &= 
    \begin{pmatrix}
        \cos \theta_0 & - \sin \theta_0 \\
        \sin \theta_0 & \cos \theta_0 
    \end{pmatrix}
    \begin{pmatrix}
        \lambda_1 & \\
        & \lambda_2 
    \end{pmatrix}
    \begin{pmatrix}
        \cos \theta_0 & \sin \theta_0 \\
        -\sin \theta_0 & \cos \theta_0 
    \end{pmatrix}\\
    &= 
    (1-\kappa)
    \begin{pmatrix}
        1 & 0 \\
        0 & 1
    \end{pmatrix}
    +\gamma
    \begin{pmatrix}
        \cos 2\theta_0 & \sin 2\theta_0 \\
        \sin 2\theta_0 & -\cos 2\theta_0
    \end{pmatrix}
\end{align}
where
\begin{align}
    \kappa &= 1 - \frac{\lambda_1 + \lambda_2}{2} \\
    \gamma &= \frac{\lambda_1 - \lambda_2}{2}.
\end{align}
are respectively the convergence and the shear.

\subsection{Curved Arc Basis}
In the case when $\theta_n(t) = a_1 t$, $\lambda_1(t) = \text{const}$, $\lambda_2(t) = \text{const}$, $F(t) = 0$, and $H_0 = 0$ the lens model reduces to a curved arc basis \citep{curved_arc_basis}. To make this easier to read we will also set $\theta_0 = 0$ as we can always choose our basis vectors in $\bm x$ such that this is true. Our model gives:
\begin{align}
    \bm{\tilde y}\left[
    \begin{pmatrix}
    t\\
    s
    \end{pmatrix}
    \right]
    &=
    \frac{\lambda_1}{a_1}
    \begin{pmatrix}
        \sin a_1 t \\
        1 - \cos a_1 t
    \end{pmatrix}
    +
    s\lambda_2
    \begin{pmatrix}
        -\sin(a_1 t) \\
        \cos(a_1 t)
    \end{pmatrix} \\
    &= 
    \frac{\lambda_1}{a_1}
    \begin{pmatrix}
        0 \\
        1
    \end{pmatrix}
    +
    \left(
    \frac{\lambda_1}{a_1}
    - s\lambda_2
    \right)
    \begin{pmatrix}
        \sin(a_1 t)\\
        -\cos(a_1 t)
    \end{pmatrix}
    \label{eq:eigen_reduced_to_cab}
\end{align}
also
\begin{align}
    \begin{pmatrix}
        x_1 \\
        x_2
    \end{pmatrix}
    &=
    \begin{pmatrix}
        \frac{1}{a_1}\sin(a_1 t) - s \sin(a_1 t) \\
        \frac{1}{a_1}(1-\cos(a_1 t)) + s\cos(a_1 t)
    \end{pmatrix} \\
    &= 
    \frac{1}{a_1}
    \begin{pmatrix}
        0\\
        1
    \end{pmatrix}
    +
    \left(\frac{1}{a_1} - s\right)
    \begin{pmatrix}
        \sin(a_1 t)\\
        -\cos(a_1 t)
    \end{pmatrix}
    \label{eq:x1x2_ts}
\end{align}
Curved arc basis lens model can be written as (derived from Eq. (27) in \cite{curved_arc_basis}):
\begin{align}
    \bm{y}_\mathrm{CAB}\left[
    \begin{pmatrix}
        x_1 \\
        x_2
    \end{pmatrix}
    \right] &=
    B
    \begin{pmatrix}
        x_1 \\
        x_2
    \end{pmatrix}
    -
    \frac{A}{\left(x_1^2 + (x_2 - r)^2\right)^{1/2}}
    \begin{pmatrix}
        x_1 \\
        x_2 - r
    \end{pmatrix}
    -
    A
    \begin{pmatrix}
        0\\
        1
    \end{pmatrix}.
\end{align}
We can write $x_1$ and $x_2$ in terms of $t$ and $s$ using Eq. \eqref{eq:x1x2_ts}, while also setting $r = 1/a_1$, $B = \lambda_2$ and $A = (\lambda_1 + \lambda_2)/a_1$
\begin{align}
    \bm{y}_\mathrm{CAB}\left[
    \begin{pmatrix}
        x_1 \\
        x_2
    \end{pmatrix}
    \right] &=
    \frac{B}{a_1}
    \begin{pmatrix}
        0\\
        1
    \end{pmatrix}
    +
    B
    \left(\frac{1}{a_1} - s\right)
    \begin{pmatrix}
        \sin(a_1 t)\\
        -\cos(a_1 t)
    \end{pmatrix} \\
    &\quad - A
    \begin{pmatrix}
        \sin(a_1 t)\\
        -\cos(a_1 t)
    \end{pmatrix}
    - A
    \begin{pmatrix}
        0\\
        1
    \end{pmatrix} \\
    &=\left(\frac{B}{a_1} - A\right)
    \begin{pmatrix}
        0\\
        1
    \end{pmatrix}
    +
    \left(\frac{B}{a_1} - Bs - A\right)
    \begin{pmatrix}
        \sin(a_1 t)\\
        -\cos(a_1 t)
    \end{pmatrix} \\
    &=   
    \frac{\lambda_1}{a_1}
    \begin{pmatrix}
        0 \\
        1
    \end{pmatrix}
    +
    \left(
    \frac{\lambda_1}{a_1}
    - s\lambda_2
    \right)
    \begin{pmatrix}
        \sin(a_1 t)\\
        -\cos(a_1 t)
    \end{pmatrix}
\end{align}
which is the same expression as Eq. \eqref{eq:eigen_reduced_to_cab}.

%
%

\end{document}